\documentclass[conference]{IEEEtran}
\IEEEoverridecommandlockouts
\usepackage{amsmath,amssymb,amsfonts}
\usepackage{algorithm}
\usepackage[noend]{algorithmic}
\usepackage{textcomp}
\usepackage{xcolor}    
\usepackage{wasysym} 
\usepackage{amsthm}  
\usepackage{graphicx,times,cite, epsfig, array, mathrsfs}  
\usepackage{array}
\usepackage{multirow}  
\usepackage[caption=false,font=scriptsize,labelfont=sf,textfont=sf]{subfig}
\usepackage{textcomp}
\usepackage{stfloats}
\usepackage{url}
\usepackage{verbatim}
\usepackage{threeparttable}
\usepackage{cases}
\usepackage{flushend}

\usepackage{cite}
\usepackage[all=normal, paragraphs=tight, floats=tight, mathspacing=tight]{savetrees}

\newtheorem{definition}{Definition}
\newtheorem{theory}{Theorem}
\newtheorem{lemma}{Lemma}
\newtheorem{remark}{Remark}
\newtheorem{corollary}{Corollary}

\begin{document}

\title{
Dynamic Cooperative MAC Optimization in RSU-Enhanced VANETs: A Distributed Approach
\vspace{-0.8cm}}
\author{ 
\IEEEauthorblockN{
Zhou~Zhang\IEEEauthorrefmark{1},
Saman~Atapattu\IEEEauthorrefmark{2},
Yizhu~Wang\IEEEauthorrefmark{1},  
Sumei~Sun\IEEEauthorrefmark{3}, 
and  Kandeepan~Sithamparanathan\IEEEauthorrefmark{2}}
\vspace{-0.4cm}
\\
 \IEEEauthorblockA{
 \IEEEauthorrefmark{1}
 National Innovation Institute of Defense Technology, Beijing, China.
 \\
 \IEEEauthorrefmark{2}School of Engineering, RMIT University, Melbourne, Victoria, Australia. \\
 \IEEEauthorrefmark{3}Institute for Infocomm
Research, Agency for Science, Technology and Research, Singapore. \\
\IEEEauthorblockA{
\IEEEauthorrefmark{1}\{zt.sy1986, wangyizhuj\}@163.com;
\IEEEauthorrefmark{2}\{saman.atapattu,kandeepan.sithamparanathan\}@rmit.edu.au
;
\IEEEauthorrefmark{3}sunsm@i2r.a-star.edu.sg
}
}
}

\maketitle

\begin{abstract}
This paper presents an optimization approach for cooperative Medium Access Control (MAC) techniques in Vehicular Ad Hoc Networks (VANETs) equipped with  Roadside Unit (RSU) to enhance network throughput. 
Our method employs a distributed cooperative MAC scheme based on Carrier Sense Multiple Access with Collision Avoidance (CSMA/CA) protocol, featuring selective RSU probing and adaptive transmission. It utilizes a dual timescale channel access framework, with a ``large-scale'' phase accounting for gradual changes in vehicle locations and a ``small-scale'' phase adapting to rapid channel fluctuations.
We propose the RSU Probing and Cooperative Access (RPCA) strategy, a two-stage approach based on dynamic inter-vehicle distances from the RSU. Using optimal sequential planned decision theory, we rigorously prove its optimality in maximizing average system throughput per large-scale phase.
For practical implementation in VANETs, we develop a distributed MAC algorithm with periodic location updates. It adjusts thresholds based on inter-vehicle and vehicle-RSU distances during the large-scale phase and accesses channels following the RPCA strategy with updated thresholds during the small-scale phase. 
Simulation results confirm the effectiveness and efficiency of our algorithm. 
\end{abstract}

\vspace{-0.2cm}
\section{Introduction} 
\label{s:intro}
\vspace{-0.1cm}
Swift advances in computing, vehicular infrastructure, and automated vehicle tech boost Vehicular Ad Hoc Networks (VANETs)\cite{Qi2021}. Research in vehicular cooperative communications addresses safety, traffic monitoring, and intersection management\cite{Gho2022, Duan2020}. Roadside units (RSUs) are crucial nodes in VANETs~\cite{Byungjin2021}. Integrating cooperative Medium Access Control (MAC) with RSUs enhances network performance via spatial diversity in Vehicle-to-Vehicle (V2V) and Vehicle-to-RSU (V2R) links~\cite{ Byungjin2021, Nguyen2022,Pari2023}. Challenges include autonomy, changing channels, and network mobility. 
Extensive research on VANET scenarios with RSUs has explored strategies for RSU-cellular V2V and RSU-relaying V2V networks, aiming to optimize system performance and address various network metrics, including throughput, error rates, outage probability, and security~\cite{Nguyen2022, Byungjin2021, Pari2023,Wei2020acm, Kimura2021}.
Centralized MAC strategies
utilize RSUs as central nodes to coordinate data transmission in V2V and V2R links, with a focus on maximizing network throughput and minimizing packet error probabilities~\cite{ Nguyen2022, Byungjin2021,Pari2023}. 
To address limitations of centralized MAC strategies, distributed cooperative MAC strategies with RSU-relaying MAC protocols have been proposed~\cite{atapattu2019physical,Wei2020acm, Kimura2021}, utilizing RSUs as relay nodes in ad hoc V2V communication. Some approaches focus on MAC mechanism design based on static user locations, while others concentrate on performance analysis using random spatial distribution models, \cite{Inaltekin2021,fang2022optimum}, with traditional MAC protocols~\cite{Kimura2021,xie2023optimal}.
\begin{figure}[!t]
\centering
{\includegraphics[width=.73\linewidth]{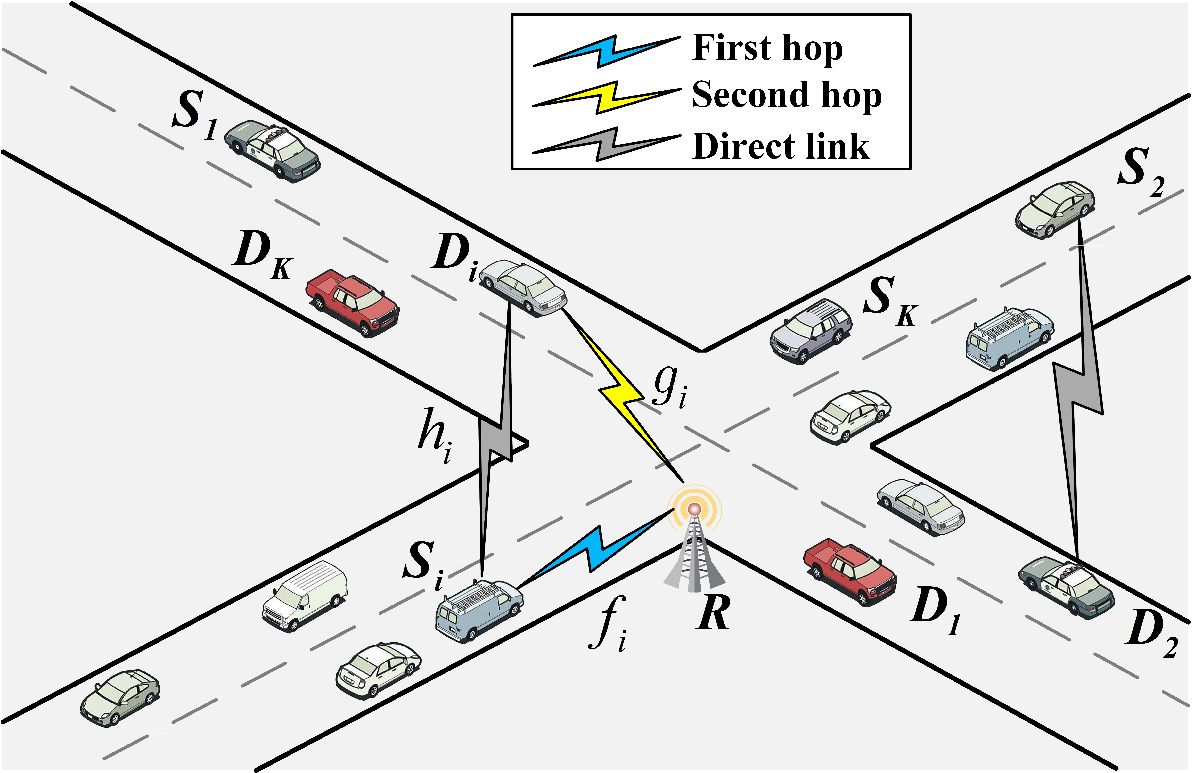}}
\vspace{-1mm}
\caption{Topology of a VANET setup with a roadside unit (RSU)}
\label{f:system_mod}
 \vspace{-6mm}
\end{figure}
Current research on cooperative MAC with RSUs for VANETs is in its early stages, posing challenges related to spatial and time diversity, including utilizing diversity from cooperative vehicles and RSUs and addressing fast-changing channel conditions and vehicular mobility. Addressing these challenges requires a collaborative MAC scheme to manage channel contention, RSU acquisition, and RSU-assisted channel access.
The distributed and mobile nature of the  network requires low-complexity, location-based operations for implementations. 
{\it This work introduces a distributed cooperative MAC solution for VANETs that leverages RSUs and establishes a robust statistical optimization framework}. Its contributions to existing research are as follows:
\begin{enumerate}
	\item {\it Selective RSU probing and transmission scheme for VANET}: We propose a new distributed cooperative MAC scheme for multiple vehicle pairs, extending CSMA/CA within a two-timescale channel access framework to handle dynamic channel and vehicle location changes.
 \item {\it Optimal cooperative MAC strategy}: We develop an optimal strategy, called RPCA, to maximize network throughput by converting a sequential planned decision optimization problem into a threshold decision strategy.
\item {\it Location-updates based cooperative MAC algorithm}: Leveraging dual timescales, we design an efficient distributed MAC algorithm capitalizing on location change-based strategy updates' higher complexity and wireless channel variations' lower complexity. Via simulations confirm our strategy's effectiveness and superiority.
 
\end{enumerate} 

 


\section{System Model and Problem Formulation}\label{sub:system_model}
\subsection{Network Model}\label{sub:network_model}
\begin{figure}[ht!]
	\begin{center}
		\includegraphics[scale=.42]{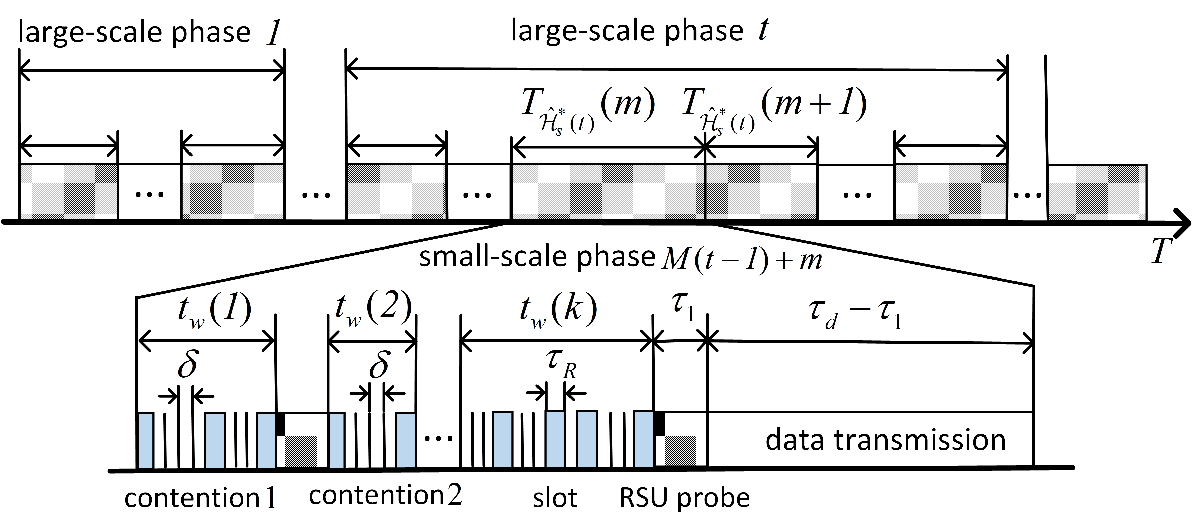}\vspace{-2mm}
		\caption{{{Time framework in dual timescales.}}}\label{f:time_scale1}
	\end{center}
  \vspace{-5mm}
\end{figure}
In Fig.~\ref{f:system_mod}, we depict a VANET with an RSU using the amplify-and-forward (AF) relaying protocol. The network comprises $K$ V2V communication pairs, each consisting of a source $S_i$ and its corresponding destination $D_i$, where $i=1,\ldots,K$. Source and destination locations are denoted as $\mathbf{x}_{S,i} \in \mathbb{R}^2$ and $\mathbf{x}_{D,i} \in \mathbb{R}^2$, respectively. 
Wireless channel characteristics, dependent on point-to-point distances, exhibit temporal variations. An RSU, denoted as $R$ with location $\mathbf{x}_{R} \in \mathbb{R}^2$, is deployed in a fixed position. The transmission powers for sources and the RSU are $P_s$ and $P_r$, respectively. In the network, channel gains for V2V links ($S_i$ to $D_i$), source vehicle-to-RSU links ($S_i$ to $R$), and RSU-to-destination vehicle links ($R$ to $D_i$) are represented as $h_i$, $f_{i}$, and $g_i$, respectively. All wireless channels exhibit independent complex Gaussian fading. 
Noise values at the RSU and destinations are independent additive white Gaussian noise (AWGN) with power $N_0$.
Consequently, the received signal-to-noise ratios (SNRs) at $D_i$ via the V2V link: $\gamma_i = P_s |h_i|^2/N_0$; at $R$: ${\gamma_{1,i}}= P_s |f_{i}|^2/N_0$; and at $D_i$ of the RSU  link: ${\gamma_{2,i}}= P_r |g_{i}|^2/N_0$. 
 The SNRs $\gamma_i$, $\gamma_{1,i}$, and $\gamma_{2,i}$  follow independent exponential distributions with mean values
{$P_{s}||\mathbf{x}_{S,i}-\mathbf{x}_{D,i}||_2^{-\alpha_1}/N_0$,
$P_{s}||\mathbf{x}_{S,i}-\mathbf{x}_{R}||_2^{-\alpha_2}/N_0$, and 
$P_{r}||\mathbf{x}_{R}-\mathbf{x}_{D,i}||_2^{-\alpha_2}/N_0$,} respectively. The path loss exponents for the V2V link and links between vehicles and the RSU are denoted as $\alpha_1$ and $\alpha_2$, respectively. Additionally, all channels experience fast fading, and the maximum transmission duration for each vehicle, denoted as $\tau_d$, is determined by the average relative motion between vehicles and the RSU.

We split channel access into two phases: a ``large-scale'' phase for gradual location changes and a ``small-scale'' phase for rapid channel fluctuations~\cite{Inaltekin2021}. The dual timescale framework is depicted in Fig.~\ref{f:time_scale1}.
Each large-scale phase includes multiple small-scale phases denoted as $m=1,\cdots,M$. Within each large-scale phase, we analyze the MAC problem considering a snapshot of vehicle locations. 

\vspace{-0.2cm}
\subsection{Cooperative RSU-assisted V2V MAC Process} \label{sub:ch_ass}
\vspace{-0.1cm}
As shown in Fig.~\ref{f:time_scale1}, within each large-scale phase, during the $m$th small-scale phase, the VANET employs a distributed wireless channel access approach using CSMA/CA protocol, with RTS/CTS mechanism \cite{ Wei2020acm, Loginov2022}. In each time slot, sources independently contend for the channel, sending RTS packets with a probability of $p_0$ and including location information. Three scenarios are considered: 
i) If no source transmits an RTS, the channel remains {\it idle}, and contention continues in the next slot.
ii) If multiple sources transmit RTS packets, a {\it collision} occurs, and contention continues in the next slot.
iii) If only one source, i.e., wining source $S_w$, sends an RTS, it {\it wins} the contention. Upon receiving the RTS, $S_w$'s destination $D_w$ and the RSU obtain CSI from $S_w$. Then, $D_w$ makes one of three decisions in the first stage:
    \begin{enumerate}
        \item \texttt{Stop}: $D_w$ sends a CTS, allowing $S_w$ to transmit at the highest achievable rate in the V2V link.
        \item \texttt{Re-contend}: $D_w$ gives up channel access, and  sends a CTS. Then, $S_w$ re-contends with other sources.
        \item \texttt{Probe RSU}: 
        $D_w$ sends a CTS to $S_w$ and the RSU, allowing $D_w$ to obtain instantaneous CSI of the V2R links, and 
        then decides to \texttt{stop} or \texttt{re-contend} in the second stage. If \texttt{stop} is chosen, $D_w$ sends a CTS, enabling $S_w$ to transmit under the RSU relaying scheme. Otherwise, $D_w$ gives up transmission, and other sources can detect an idle slot after RTS/CTS exchange, indicating \texttt{re-contend} is chosen, initiating a new contention among all sources.
    \end{enumerate}
\vspace{-2mm}
\subsection{Problem Formulation}\label{sub:prob_form}\vspace{-2mm}
For cooperative access, we employ an optimal sequential planned decision theory~\cite{Schmitz_N_book}, using a three-step approach. 
%

\subsubsection{Sequential observation process}
We establish sequential observation process with two observations: the  {\it First Obs.} and the  {\it Second Obs.} After the $k$th successful channel contention: 

\begin{itemize}
	\item  {\it First Obs. (Obs.~($n=2k-1$))} involves channel contention among sources until a winning pair emerges. In the $k$th successful contention, $w(k)$ is the $S_w$ index, and $t_w(k)$ is the contention time, following a geometric distribution with a parameter of $Kp_0(1-p_0)^{K-1}$~\cite{Wei2020acm, Loginov2022}. The last contention slot in an observation is successful, lasting $\tau_{R}+\tau_{C}$, where $\tau_{R}$ and $\tau_{C}$ are RTS and CTS durations. Any other contention results in either an {\it idle} slot or a {\it collision}. 
 The mean observation duration is: 
 $\tau_o= \tau_{R}+ \tau_{C}+\frac{(1-p_0)^{K}\delta}{Kp_0(1-p_0)^{K-1}}+\frac{(1-(1-p_0)^{K}-Kp_0(1-p_0)^{K-1})\tau_{R}}{Kp_0(1-p_0)^{K-1}}$.
After a successful  contention,  $D_w$ acquires the  SNR $\gamma_{w(k)}(k)$ for the V2V link and decides whether to \texttt{stop}. If \texttt{stop} is chosen, $S_w$ transmits data in the V2V link; if \texttt{probe RSU} is chosen, $S_w$ selects the RSU for the \textit{Second Obs.}; otherwise, the source \texttt{re-contends}. 

\item {\it Second Obs. (Obs.~($n=2k$))} is an optional RSU probing by the winning pair. The destination probes the RSU for a duration $\tau_1=\tau_{R}+\tau_{C}$. After obtaining RSU channel gains $f_{w(k)}(k),g_{w(k)}(k)$, it calculates the maximal achievable rate and decides whether to \texttt{stop}. If \texttt{stop} is chosen, it utilizes either the V2V (direct) transmission or the RSU-aided transmission, selecting the highest rate, as in \eqref{e:reward_definition}.

\end{itemize}

Considering that RSU probing and access decisions rely on sequential observations, we can depict the ``observation path'' for strategy $\mathcal{H}$ up to the $k$th successful channel contention as:
\vspace{-1mm}
\begin{equation}\label{e:path_definition}
\begin{aligned}
\mathcal{H} = \begin{cases}
    (1,{R}_1,1,{R}_2,...,1), & \text{odd } {|\mathcal{H}|}  \\
    (1,{R}_1,1,{R}_2,...,1,{R}_k), & \text{even }{|\mathcal{H}|} 
\end{cases}
\end{aligned}
\vspace{-1mm}
\end{equation}
where {${R}_1,\cdots,{R}_k\in\{0,1\}$}. 
Along the observation path, the element $1$ signifies that after a successful channel contention, a source wins the channel and acquires CSI of the V2V link. The element ${R}_k$ represents the winner pair's decision regarding RSU probing. Specifically, ${R}_k=1$ indicates RSU probing, while ${R}_k=0$ signifies otherwise.

\subsubsection{Reward and objective function}
For the $k$th successful channel contention, the instantaneous reward $Y_{\mathcal{H}}(t)$ and time cost $T_{\mathcal{H}}(t)$ by {\tt stop} are defined based on observation path ${\mathcal{H}}$. Further, $Y_{\mathcal{H}}$ represents the maximum transmitted data quantity in bits and can be expressed in different ways depending on the observation path:
\begin{equation}\label{e:reward_definition}
\begin{aligned}
Y_{\mathcal{H}}= \begin{cases}
    \tau_d R_{d,k}, & \text{odd } |{\mathcal{H}}| \\
    \tau_{d,1}\max\{R_{d,k},R_{r,k}\},\,&\, \text{even } |{\mathcal{H}}| \& {R}_{k}=1
\end{cases}
\end{aligned}
\vspace{-1mm}
\end{equation}
In particular, the achievable rates by V2V and RSU-aided transmissions are represented as $R_{d,k}=\log_2\big(1+\gamma_{w(k)}(k)\big)$ and $R_{r,k}=(1/2)\log_2\big(1+\gamma_{w(k),r}(k)\big)$, respectively, where received SNR at $D_w$ is given as $\gamma_{w(k),r}(k)=\gamma_{w(k)}+\frac{{\gamma_{1,w(k)}}{\gamma_{2,w(k)}}}{{\gamma_{1,w(k)}}+{\gamma_{2,w(k)}}+1}$. The duration of data transmission time is given by {$\tau_{d,1}=\tau_d-\tau_1$}. 
Moreover, the time cost $T_{\mathcal{H}}$ represents the total time spent until the $k$th channel contention, including the data transmission duration. It can be expressed as
	$T_{\mathcal{H}}{=}\sum_{l=1}^k t_w(l) +\sum_{l=1}^{k-1}\mathbb{I}[{R}_l=1]\tau_1+\tau_d$.
If the observation path $\mathcal{H}$ leads to the decision to \texttt{stop}, denoted as $\mathcal{H}_s=\mathcal{H}$, the instantaneous throughput $Y_{\mathcal{H}}/T_{\mathcal{H}}$\,{[bits/sec]} is achieved through data transmission. Otherwise, the observation continues with an updated path $\mathcal{H}$.
Since $\mathcal{H}_s$ encompasses the observation path and stop decision, it can articulate a strategy for the V2V distributed access outlined in Section~\ref{sub:ch_ass}.


\subsubsection{Statistical optimization}
Recognizing the stochastic nature of the multiple V2V channel access process, strategy $\mathcal{H}_s$ is repeated for $M$ {small-scale} phases,
and the network throughput
can be expressed as $\mathbb{E}[Y_{\mathcal{H}_s}]/\mathbb{E}[T_{\mathcal{H}_s}]$. 
Our goal is to find the optimal strategy $\mathcal{H}_s^*$  from a set of feasible channel access strategies $\mathcal{H}_s$,  referred as {\it Optimal RSU Probing and Cooperative Access (RPCA)}, aiming to maximize the average network throughput
$\lambda^*$ in bits/sec, given by
\begin{equation}\label{ts:problem}
	\mathcal{H}_s^*=\arg\;\;\sup_{\mathcal{H}_s}\mathbb{E}[Y_{\mathcal{H}_s}]/\mathbb{E}[T_{\mathcal{H}_s}]\, \text{ and } \,\lambda^*={\mathbb{E}[{Y_{\mathcal{H}_s^*}}}]/{\mathbb{E}[{T_{\mathcal{H}_s^*}}]}.
\end{equation}
 \vspace{-3mm}
 
\section{Optimal RPCA Strategy and Algorithm}\label{s:probem_solution}


\vspace{-0.1cm}
\subsection{General Formulation of the Optimal Strategy}
As in \cite[Theorem~1]{fergusonoptimal},
optimization of  $\mathbb{E}[Y_{\mathcal{H}_s}]/\mathbb{E}[T_{\mathcal{H}_s}]$ is equivalent to optimal sequential planned decision (SPD) problem maximizing the 
reward $\sup_{{\mathcal H}_s}\big\{\mathbb{E}[Y_{{\mathcal H}_s}]-\lambda \mathbb{E}[T_{{\mathcal H}_s}]\big\}$ for a price $\lambda>0$.
Therefore, we first find an optimal strategy ${\mathcal H}_s^*(\lambda)$ by solving problem $\sup_{{\mathcal H}_s}\big\{\mathbb{E}[Y_{{\mathcal H}_s}]-\lambda\mathbb{E}[T_{{\mathcal H}_s}]\big\}=0$, and then derive the optimal strategy ${\mathcal H}_s^*={\mathcal H}_s^*(\lambda^*)$ where $\sup_{{\mathcal H}_s}\big\{\mathbb{E}[Y_{{\mathcal H}_s}]-\lambda^*\mathbb{E}[T_{{\mathcal H}_s}]\big\}=0$.
We determine the optimal ${\mathcal H}_s^*(\lambda)$ using the principle of optimality. The comparison thresholds for instant reward are defined as follows. 



\begin{definition}\label{def:w_func}
 For 
 price $\lambda>0$, we define threshold function $W_{i}(\gamma,\lambda)$ for RSU probing  with respect to pair $i=1,\cdots,K$ as $W_{i}(\gamma,\lambda):=\mathbb{E}\big[\max \big\{\tau_{d,1}\cdot\max\{R_{d,k},R_{r,k}\}-\lambda\tau_d,{-\lambda\tau_1}\big\}\big|w(k)=i,\gamma_{i}(k)=\gamma \big]$.
 It represents the maximum expected reward of $\{Y_{{\mathcal H}_s}-\lambda T_{{\mathcal H}_s}\}$, achieved by probing the RSU while following an optimal strategy for subsequent MAC processes, given the $S_w$ as $w(k)=i$ and V2V link $\gamma_{i}(k)=\gamma$. The time cost of prior observations, denoted as $\sum_{l=1}^k t_w(l) +\sum_{l=1}^{k-1}\mathbb{I}[{R}_l=1]\tau_1$, is excluded. The reward is computed by selecting the larger value between the instant reward from RSU-assisted transmission and the anticipated future reward. 
\end{definition}
Due to the i.i.d. statistical properties of winner pair CSI and contention time, the value is equal for all $k\geq 1$.
Using the optimal SPD theory~\cite{Schmitz_N_book}, we derive the optimal  ${\mathcal H}_s^*$ in a general form.
\begin{theory}\label{th:optimal_rule1}
An optimal strategy $\mathcal{H}_s^*$ for maximizing $\sup_{\mathcal{H}_s}\{\mathbb{E}[Y_{\mathcal{H}_s}]/\mathbb{E}[T_{\mathcal{H}_s}]\}$ is as follows:
Start with $k=1$, and then continue observing until the condition {\tt stop} is met. {Specially,} after the $k$th successful channel contention, 
\begin{enumerate}
    \item At Obs.~$(2k-1)$:
	\begin{itemize}
		\item  If the immediate reward $\tau_d R_{d,k}-\lambda^* \tau_d\ge \max\big\{W_{w(k)}\big(\gamma_{w(k)}(k),\lambda^*\big),0\big\}$, then {\tt stop} and transmit over the V2V link.
			\item If the immediate reward $\tau_d R_{d,k}-\lambda^* \tau_d<0$ and $W_{w(k)}\big(\gamma_{w(k)}(k),\lambda^*\big)<0$, then {\tt re-contend} by proceeding to ($k+1$)th channel contention.
		\item Otherwise, {\tt probe RSU}.
	\end{itemize}	
	\item At Obs.~($2k$):
		\begin{itemize}
			\item {If the immediate reward $\max\big\{R_{d,k},R_{r,k}\big\}\ge \lambda^*$ where  $R_{r,k}=(1/2)\log_2\big(1+\gamma_{w(k),r}
   \big)$}, then {\tt stop}.
			\item Otherwise, {\tt re-contend} by proceeding to the $(k+1)$th channel contention.
		\end{itemize}
 \end{enumerate}
The maximal average network throughput $\lambda^*$ uniquely exists, and is determined by the solution of
\begin{align}
	&\mathbb{E}\big[
	\max\big\{\!\tau_d R_{d,k}-\lambda^*\tau_d,0,W_{w(k)}\big(\gamma_{w(k)}(k),\lambda^*\big)
	\big\}\big]\!=\!
	\lambda^*\tau_o.\label{equ:bellman1}
\end{align}
\end{theory}
\begin{IEEEproof}
 This can be proved by Theorem 2.14 in \cite{Schmitz_N_book}.
\end{IEEEproof}

\begin{remark}
Theorem~\ref{th:optimal_rule1} establishes the optimal strategy $\mathcal{H}_s^*$ using sequential planned decision theory. It demonstrates a two-stage decision strategy based on multiple channel contentions and observed instantaneous CSI of V2V and V2R links. To ensure feasibility, two steps are necessary: 1) Finding the maximal average network throughput $\lambda^*$ by solving (\ref{equ:bellman1}) offline; 2) Deriving analytical expressions for $W_{w(k)}\big(\gamma_{w(k)}(k),\lambda^*\big)$. 
Moreover, the analysis of (\ref{equ:bellman1}) involves deriving analytical expressions of $W_{i}\big(\gamma,\lambda\big)$ for $i=1,\cdots,K$ 
and a given $\lambda$. 
\end{remark}
In order to solve (\ref{equ:bellman1}) and reduce the complexity, we next provide expressions for
$W_{i}\big(\gamma,\lambda\big)$. 
\subsection{Closed-form Expression}


By using approximation $\frac{{\gamma_{1,i}}{\gamma_{2,i}}}{{\gamma_{1,i}}+{\gamma_{2,i}}+1}{\approx}\min\{{\gamma_{1,i}},{\gamma_{2,i}}\}$, we derive the analytical expression $\hat{W}_{i}(\gamma,\lambda){\approx} W_{i}(\gamma,\lambda)$ 
 as follows:
\begin{align*}
 & \hat{W}_{i}(\gamma,\lambda)\overset{\triangle}{=}\frac{\tau_{d,1}}{2\ln 2}
		\big(e^{-c_{i} (\Omega-\gamma)}\ln(\Omega+1)
		+ e^{c_{i}(1+\gamma)}{\rm{E}_1}\big(c_{i}(\Omega+1)\big)\big)
		\nonumber\\&+
  \tau_{d,1}\big(\mathbb{I}[R_{d,k}\!>\!\lambda]
		(R_{d,k}\!-\!\lambda) F_{r,i}(\gamma^2\!+\!\gamma)
		\!+\!\lambda   F_{r,i}( \Omega\!-\!\gamma )\big)\!-\!\lambda\tau_d
	\end{align*}
 where ${\rm{E}_1}(\cdot)$ is the exponential integral function, $\Omega=\max\{4^{\lambda}-1,\gamma^2+2\gamma\}$, $c_{i}\!=\!N_0\big(P_s^{-1}(||\mathbf{x}_{R}-\mathbf{x}_{D,i}||_2)^{\alpha_2} + P_r^{-1}(||\mathbf{x}_{S,i}-\mathbf{x}_{R}||_2)^{\alpha_2} \big)$, and $F_{r,i}(x)=1-e^{-c_{i}\cdot x}$.
The analytical expression above 
provides a means to simplify the optimal strategy $\mathcal{H}_s^*$.
To ensure the feasibility of the proposed strategy, we 
refine it into strategy $\hat{\mathcal{H}}_s^*$,
such that all decision functions can be calculated from closed-form expressions $\hat{W}_{i}\big(\gamma,\lambda\big)$. 

\vspace{-1mm}
\subsection{Pure Threshold Decision Strategy}
We analyse threshold-based criteria and fine-tune our proposed  $\hat{\mathcal{H}}_s^*$
to a two-stage decision. We thoroughly explore the decisions' conditions for  {\tt stop}, {\tt probe RSU}, and {\tt re-contend} within  strategy $\hat{\mathcal{H}}_s^*$ elucidated in Theorem~\ref{th:optimal_rule1}.

Let us define thresholds $\zeta_{i}\ge0$ for the winner pair $i$, where $i=1,\cdots,K$. These thresholds ensure that $\hat{W}_{i}^*(\gamma,\lambda^*)=0$. Additionally, let us define thresholds $\eta_{i}$ such that $\hat{W}_{i}^*(\gamma,\lambda^*)=\tau_d\left(\log_2(1+\gamma)-\lambda^*\right)$.
These thresholds act as boundaries for comparing rewards associated with optimal decisions and possess specific properties, as stated in the following lemma.

\begin{lemma}\label{l:threshold_property}
For every winner pair $i$, there exist unique thresholds $\zeta_{i}$ and $\eta_{i}$, 
satisfying the following two properties:
\begin{enumerate}
    \item If $\gamma \geq \zeta_{i}$, $\hat{W}_{i}^*(\gamma,\lambda^*) \geq 0$; If $\gamma < \zeta_{i}$, $\hat{W}_{i}^*(\gamma,\lambda^*) < 0$.
    \item If $\gamma \geq \eta_{i}$, $\hat{W}_{i}^*(\gamma,\lambda^*) \leq \tau_d\log_2(1+\gamma)-\lambda^*\tau_d$; If $\gamma < \eta_{i}$, $\hat{W}_{i}^*(\gamma,\lambda^*) > \tau_d\log_2(1+\gamma)-\lambda^*\tau_d$.
\end{enumerate}
These properties describe how the optimal threshold-based decisions behave in relation to the values of $\gamma$.
\end{lemma}
\begin{IEEEproof}
We use the monotonicity of $\hat{W}_{i}^*(\gamma,\lambda^*)$ and $\tau_d\big(\log_2(1+\gamma)-\lambda^*\big)- \hat{W}_{i}^*(\gamma,\lambda^*)$  with respect to $\gamma$.
\end{IEEEproof}	
The thresholds $\zeta_{i}$ and $\eta_{i}$ are predetermined and fixed, as in Lemma~\ref{l:threshold_property}. They can be directly compared with CSI $\gamma_{w(k)}(k)$ for decision-making.
Then, we enhance the strategy $\hat{{\mathcal H}}_s^*$ by refining it into a pure threshold decision (PTD) strategy.
\begin{theory}\label{th:optimal_rule2}
{A refined two-stage PTD strategy}, based on  $\hat{\mathcal H}_s^*$, is as follows: After the $k$th successful channel contention, 
\begin{enumerate}
\item At Obs.~($2k-1$),
\begin{itemize}
\item If the V2V link SNR $\gamma_{w(k)}(k)\!\ge\! \eta_{w(k)}$, then {\tt stop} and transmit over the V2V link.
\item If the V2V link SNR $\gamma_{w(k)}(k)\!<\! \zeta_{w(k)}$, then {\tt re-contend} 
by proceeding to ($k+1$)th channel contention. 
\item Otherwise, then {\tt probe RSU}. 
\end{itemize}
\item At Obs.~($2k$),
\begin{itemize}
\item If $\max\big\{\gamma_{w(k)}(k),\gamma_{w(k),r}(k)\big\}\ge 2^{\lambda^*}-1$, then {\tt stop}.
\item Otherwise, {\tt re-contend} by proceeding to the ($k+1$)th channel contention.
\end{itemize}
\end{enumerate}
Moreover, the maximal average throughput $\lambda^*$ can be obtained by utilizing the iterative algorithm described in Algorithm~\ref{Algorithm_1}, 
with a desired numerical accuracy denoted by $\epsilon$.
\end{theory}
\begin{IEEEproof}
 We can obtain the PTD strategy by combining results of Lemma~\ref{l:threshold_property} and Theorem~\ref{th:optimal_rule1}.
 The algorithm convergence 
 is guaranteed 
 by Proposition~1.2.3 in \cite{Berts1999}.
\end{IEEEproof}

\vspace{-2mm}
\begin{remark}
Theorem~\ref{th:optimal_rule2} presents an optimal PTD strategy, 
where each V2V pair, following a successful channel contention, determines its action (transmission, RSU probing, or re-contend channel) by comparing its instantaneous SNR with fixed thresholds. This strategy has an online complexity of at most ${\cal O}(1)$, making it easy to implement.
To ensure the feasibility of the strategy, Algorithm~\ref{Algorithm_1} is provided as an iterative algorithm to compute the value of $\lambda^*$ with a desired numerical accuracy $\epsilon$. The strategy only requires network statistics and can be computed online with low frequency or offline using pre-computed values stored in a look-up table.
\end{remark}
\vspace{-5mm}
\begin{algorithm}[h!]
		\caption{Iterative algorithm for $\lambda^*$}\label{Algorithm_1}
	\renewcommand{\algorithmicrequire}{\textbf{Input:}}
	\renewcommand{\algorithmicensure}{\textbf{Output:}}
\begin{algorithmic}[1]
	\REQUIRE{$\lambda_0=0$, $\Delta=1$, $\epsilon$, $l=0$}
  \vspace{-1mm}
 \WHILE{$\Delta\ge \epsilon$}
  \vspace{-1mm}
 \STATE $l\leftarrow l+1$
  \vspace{-1mm}
	\FOR{$i=1:K$}	
	 \STATE	$\sigma_{i}^2\leftarrow P_{s}||\mathbf{x}_{S,i}-\mathbf{x}_{D,i}||_2^{-\alpha_1}/N_0$			
			\IF{$\hat{W}_i\big(2^{\lambda_l}-1,\lambda_l\big)\!>\! 0$}
				\STATE 		$\eta_{i}\leftarrow \hat{W}_{i}(\gamma,\lambda_l)=\tau_d\big(\log_2(1+\gamma)-\lambda_l\big)$	
		\STATE $\zeta_{i}\leftarrow \hat{W}_{i}(\gamma,\lambda_l)=0$
   \STATE $\Lambda_i(\lambda_{l})\leftarrow
				\frac{1}{\sigma_{i}^2}\int_{\zeta_{i}}^{\eta_{i}}
				\hat{W}_{i}^*(x,\lambda_l) e^{\frac{-x}{\sigma_{i}^2}}dx$   
				 $+\frac{\tau_d}{\sigma_{i}^2}\int_{\eta_{i}}^\infty\!\big(\log_2(1+x)\!-\!\lambda_l\big) e^{\frac{-x}{\sigma_{i}^2}}\! dx$
    
    \ELSE		
					\STATE	$\Lambda_i(\lambda_{l})\leftarrow \frac{\tau_d}{\sigma_{i}^2} \int_{2^{\lambda_{l}}\!-\!1}^\infty \big(\log_2(1\!+\!x)-\lambda_{l}\big) e^{\frac{-x}{\sigma_{i}^2}} dx$
			\ENDIF
  \ENDFOR
  	\STATE $\Delta\leftarrow\frac{1}{K}\sum_{i=1}^K\Lambda_i(\lambda_l)- \lambda_{l} \tau_o$, update $\lambda_{l+1}\leftarrow\lambda_l+\alpha \Delta$ where step-size $\alpha$ satisfies $\epsilon\le\alpha\le \frac{2-\epsilon}{\tau_o+\tau_d}$
  \ENDWHILE
	\ENSURE{$\lambda^*\leftarrow \lambda_{l}$}
	\end{algorithmic}
 \vspace{-1mm}
\end{algorithm}	
 \vspace{-3mm}
In certain cases, the optimal strategy structure may result in a situation where threshold $\eta_{w(k)}$ for the {\tt stop} decision is smaller than threshold $\zeta_{w(k)}$ for the {\tt re-contend} decision. In such cases, the optimal strategy can be simplified as follows.

\begin{definition}
    We define the V2V pair set ${\mathscr K}^*=\big\{i\in\{1,...,K\}:\hat{W}_i^*\big(2^{\lambda^*}-1,\lambda^*\big)\!>\! 0\big\}$. This set represents a category of user pairs for which the V2V transmission can benefit from RSU-assisted transmission. 
\end{definition}

By adhering to the proposed strategy $\hat{\mathcal H}_s^*$, the two-stage thresholds structure can be simplified into a single-stage pure threshold structure for certain pairs. As a result, the online complexity becomes $\mathcal{O}(1)$, as stated in Corollary~\ref{col:optimal_rule3}.

\begin{corollary}\label{col:optimal_rule3}
In the two-stage thresholds structure, the proposed strategy $\hat{{\mathcal H}}_s^*$ can be simplified to a single-stage strategy as follows: After the $k$th successful channel contention, for the winning source $w(k)\notin{\mathscr K}^*$, the following steps are followed:
\begin{enumerate}
\item If the V2V link SNR $\gamma_{w(k)}(k)\ge 2^{\lambda^*}-1$, {\tt stop} and transmit over the V2V link.
\item Otherwise, {\tt re-contend} by proceeding to the $(k+1)$th channel contention.
\end{enumerate}
\end{corollary}
\begin{IEEEproof}
{By proving for $i\notin{\mathscr K}^*$ and $\forall \gamma\!\ge\! 0$, $\hat{W}_i^*(\gamma,\lambda^*)\le \max\big\{(\log_2(1+\gamma)-\lambda^*)\tau_d,0\big\}$, the result is derived. }
\end{IEEEproof}
So far, we discussed the optimal strategy assuming fixed thresholds and static inter-distances. 

\section{Two-timescale MAC Algorithm} \label{s:net}
We propose a self-organized MAC algorithm designed for scenarios involving dynamic vehicle movement, varying inter-distance, and changing channel conditions. 
This algorithm incorporates location updates, online strategy reconfiguration, and utilizes the MAC strategy $\hat{\mathcal H}_s^*$ from Theorem~\ref{th:optimal_rule2}. 

\vspace{-2mm}
\begin{algorithm}[h!]
		\caption{MAC algorithm by strategy ${\mathcal H}_s^*(t)$}\label{Algorithm_3}
	\renewcommand{\algorithmicrequire}{\textbf{Input:}}
	\renewcommand{\algorithmicensure}{\textbf{Output:}}
\begin{algorithmic}[1]
	\REQUIRE{${\mathcal H}_s^*(t),{\mathscr K}^*$}
	\FOR{$m=1:M$}	
  \vspace{-1mm}
	 \STATE	$k\leftarrow 1$		
   \vspace{-1mm}
		\REPEAT				
			\IF{${S}_{i}$ wins channel contention}
    \vspace{-1mm}
				\STATE $D_i\leftarrow \mathbf{x}_{S,i},\gamma_i(k)$
			 \vspace{-0.5mm}	
				\IF{$i\in {\mathscr K}^*$}
				 \vspace{-1mm}
					\IF{$\gamma_i(k)>\eta_{i}$}					
					\STATE	{\tt stop} by V2V link transmission				
					\ELSIF{$\gamma_i(k)\le \zeta_{i}$}
					\STATE {\tt re-contend} 			
					\ELSE 
      \vspace{-1mm}
					\STATE {\tt probe RSU} and calculate $\gamma_{i,r}(k)$
						\IF{$\max\big\{\!\gamma_i(k),\gamma_{i,r}(k)\!\big\}\!\ge\! 2^{\lambda^*}\!\!-\!\!1$}
							
							\IF{$\gamma_{i,r}(k)>\gamma_i(k)$}
							\STATE {\tt stop} by RSU aided transmission
							\ELSE							\STATE{\tt stop} by V2V link transmission 
							\ENDIF						
						\ELSE								%
						\STATE	{\tt re-contend} 
						\ENDIF

					\ENDIF
				\ELSE		
					\IF{$\gamma_i(k)\ge 2^{\lambda^*}-1$}
						\STATE {\tt stop} by V2V link transmission
					\ELSE
					\STATE	{\tt re-contend}
			 \vspace{-1mm}		
					\ENDIF
			\ENDIF
 \vspace{-1mm}
			\ELSE
			\STATE	$D_i\leftarrow (\mathbf{x}_{S,w(k)},\mathbf{x}_{D,w(k)})$
		\ENDIF
	\UNTIL{$m$th small-scale phase ends}
		\ENDFOR
	\ENSURE{$\mathbf{\Phi}(t+1)$}
	\end{algorithmic}
 \vspace{-1mm}
\end{algorithm}	
\vspace{-3mm}
\begin{algorithm}[h!]
		\caption{Strategy ${\hat{\mathcal H}}_s^*(t)$ update }\label{Algorithm_2}
	\renewcommand{\algorithmicrequire}{\textbf{Input:}}
	\renewcommand{\algorithmicensure}{\textbf{Output:}}
\begin{algorithmic}[1]
	\REQUIRE{$\mathbf{\Phi}(t)$}
 \STATE $\lambda^*\leftarrow \text{Look-up Table by Algorithm \ref{Algorithm_1}}$
  \STATE $\zeta_{i} \leftarrow \hat{W}_{i}^*(\gamma,\lambda^*)=0$
	\STATE  $\eta_{i} \leftarrow \tau_d\big(\log_2(1+\gamma)-\lambda^*\big)=\hat{W}_{i}^*(\gamma,\lambda^*)$
 \STATE pair set ${\mathscr K}^*\leftarrow\big\{i\in\{1,...,K\}:\hat{W}_i\big(2^{\lambda^*}-1,\lambda^*\big)\!>\! 0\big\}$.
	\ENSURE{$\hat{{\mathcal H}}_s^*(t)$, ${\mathscr K}^*$}
	\end{algorithmic}
 \vspace{-1mm}
\end{algorithm}	
\vspace{-3mm}
To account for the faster fading of the wireless channel compared to vehicular mobility, our algorithm employs a two-scale time-phased structure, as illustrated in Fig.~\ref{f:time_scale1}. Large-scale phases, denoted by index $t$, are subdivided into small-scale phases, indexed as $m=M(t-1)+1,\cdots,Mt$. The duration $\tau_d$ usually spans milliseconds, while location changes often occur within seconds, with $M$ typically  in the hundreds or even thousands.
\begin{itemize}
\item {\it {Small-scale} Phase} manages low-level channel access decisions, adapting to fast wireless channel fading changes during the interval $[T,T\!+\!T_{\hat{{\mathcal H}}_s^*(t)})$. Here, $T_{\hat{{\mathcal H}}_s^*(t)}$ signifies the duration of a successful data transmission using strategy $\hat{{\mathcal H}}_s^*(t)$ (discussed in Section \ref{s:probem_solution}).

\item {\it {Large-scale} Phase} consists of multiple {small-scale} phases, in  $\big[T,T\!+\!\sum_{m=M(t-1)+1}^{Mt} T_{\hat{{\mathcal H}}_s^*(t)}(m)\big)$, where $T_{\hat{{\mathcal H}}_s^*(t)}(m)$ is duration of the $m$th {small-scale} phase. 
    
\end{itemize}
From $T=0$, each vehicle begins phases indexed at $t=1$, $m=1$, utilizing cooperative MAC strategy $\hat{{\mathcal H}}_s^*(1)$ with a simple CSMA/CA protocol, ensuring each vehicle accesses and transmits on the V2V link. In every large-scale phase, each vehicle performs three procedures.
\vspace{-2mm}
\subsection{Location Information Update Procedure}\label{sub:procedure1}\vspace{-1mm}
Using CSMA/CA with sender location in RTS/CTS, here is how distance information $\mathbf{\Phi}(t)$ updates. In the ``large-scale'' phase $t,$ each V2V pair $S_i-D_i$ updates its location info via RTS and CTS messages (Algorithm~\ref{Algorithm_3}, line 5). Additionally, when receiving RTS and CTS from other winners $S_j-D_j$ ($j\neq i$) (Algorithm~\ref{Algorithm_3}, line 26), pair $S_i-D_i$ obtains their location $(\mathbf{x}_{S,j},\mathbf{x}_{D,j}).$ At phase $t$ end, using RSU location $\mathbf{x}_{R},$ each vehicle acquires distance info,
$\mathbf{\Phi}(t)=\big(\mathbf{\Phi}_0(t),\mathbf{\Phi}_1(t),\cdots,\mathbf{\Phi}_K(t)\big)$, where $\mathbf{\Phi}_0(t)=(||\mathbf{x}_{S,1}-\mathbf{x}_{D,1}||_2,...,||\mathbf{x}_{S,K}-\mathbf{x}_{D,K}||_2)$, $\mathbf{\Phi}_i(t)=(||\mathbf{x}_{S,i}-\mathbf{x}_{R}||_2,||\mathbf{x}_{R}-\mathbf{x}_{D,i}||_2)$ with $i=1,...,K$, respectively. The periodically updated distance information $\mathbf{\Phi}(t)$ serves as input to Algorithm~\ref{Algorithm_1}.
\vspace{-2mm}
\subsection{Proposed MAC Strategy Reconfiguration}\label{sub:procedure2}
\vspace{-1mm}
At the start of large-scale phase $t$, each vehicle uses Algorithm~\ref{Algorithm_2} to reconfigure the MAC strategy based on distance information $\mathbf{\Phi}(t)$. This involves obtaining $\lambda^*$ from a precomputed table, calculating SNR-based thresholds $\{\zeta_{i},\eta_{i}\}$ (for $i=1,\cdots,K$), and determining the set ${\mathscr K}^*$. These values collectively determine the reconfigured MAC strategy $\hat{{\mathcal H}}_s^*(t)$.
\vspace{-1mm}
\subsection{Distributed RSU Assisted MAC Algorithm for VANETs}\label{sub:procedure3}
Using MAC strategy $\hat{{\mathcal H}}_s^*(t),$ each V2V pair employs distributed channel access during large-scale phase $t$ following Algorithm~\ref{Algorithm_3}. This algorithm uses the input set ${\mathscr K}^*$ and leverages results from Theorem~\ref{th:optimal_rule2} and Corollary~\ref{col:optimal_rule3}. As ${\mathscr K}^*$ categorizes pairs into two-stage and single-stage threshold structures, the proposed algorithm offers lower complexity than $\hat{{\mathcal H}}_s^*(t).$ After completing $M$ small-scale phases, the $t$th large-scale phase ends, and the process restarts with the beginning of the $(t+1)$th phase.

\vspace{-2mm}
\section{Numerical Results}\label{s:simu}

We consider $K=8$, $\beta_0=-30$\,dB, $P_s=P_r$,  $N_0=-90$\,dBm, $p_0=0.3$, $\delta=50\,\mu$s, $\tau_{R}=\tau_{C}=100\,\mu$s, path loss exponent $\alpha_{1}=3$ and $\alpha_{2}=2.5$.
We model two intersecting roads on a $2D$  system. We conduct simulations for a mobility scenario where users randomly move along the roads, adhering to a random way-point vehicle mobility model in \cite{book_mobility}. The simulation area spans $1\times1$\,km$^2$, and vehicles' speeds range in $20-80$\,km/h.
The simulated network operates with a large-scale phase number of $100$, with each phase comprising $M=300$ small-scale phases. The initial locations are set using the static scenario.
\vspace{0mm}
\begin{figure*}[!t]
\centering
\subfloat[Throughput vs $P_s$. 
\label{f:compared_b3}]
{\includegraphics[width=0.3\linewidth]{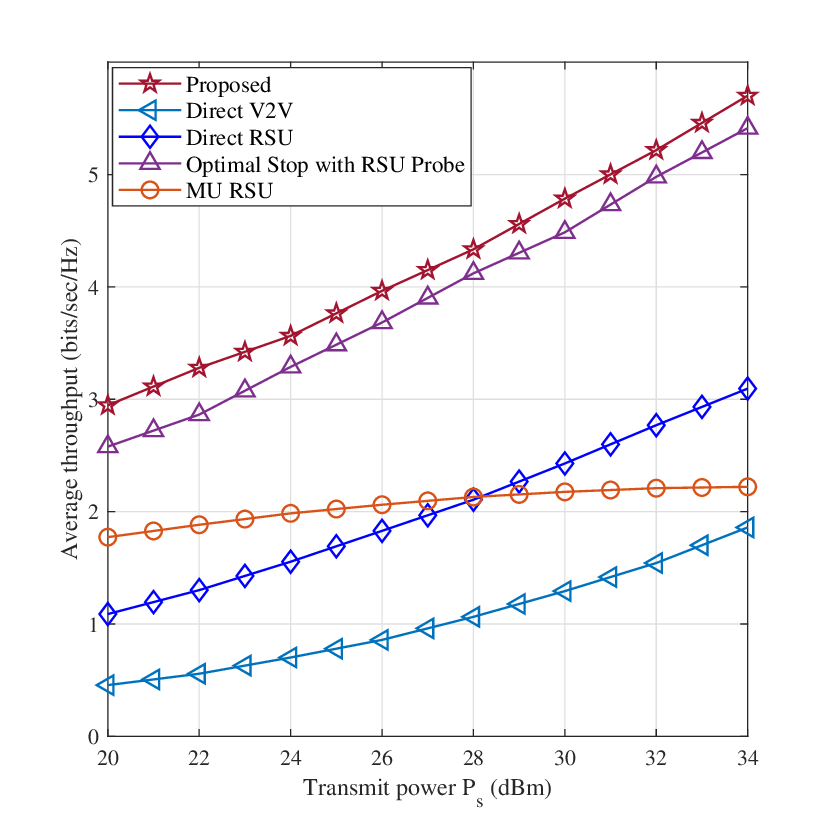}}
\hfil
\subfloat[Throughput vs $\tau_d$. 
\label{f:move_taud_compare}]
{\includegraphics[width=0.3\linewidth]{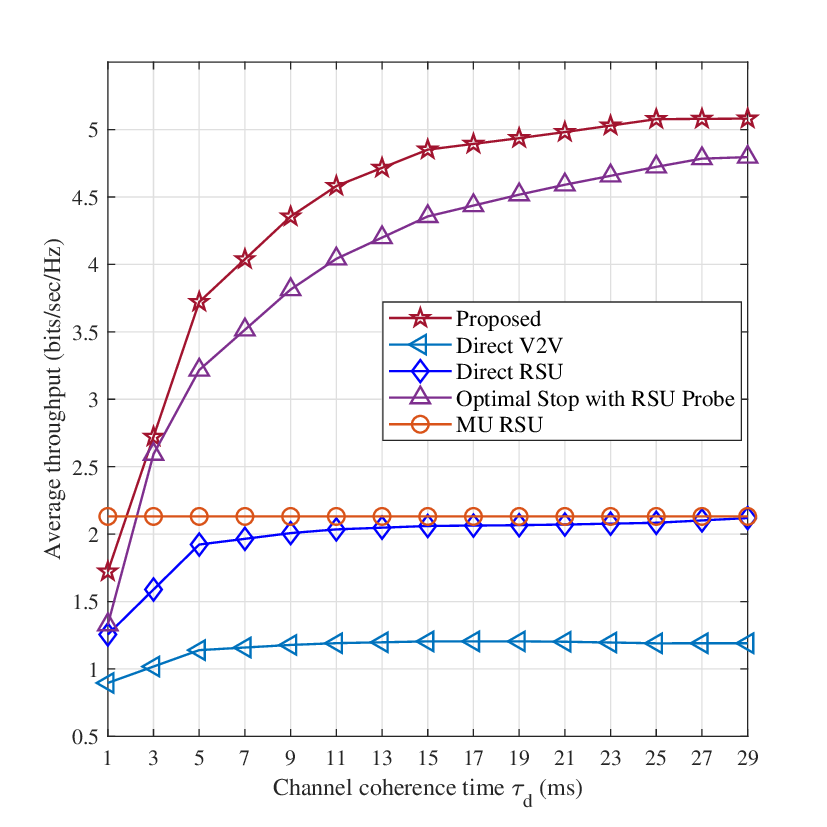}}
\subfloat[Throughput vs $p_0$. 
\label{f:compare_p0}]
{\includegraphics[width=0.3\linewidth]{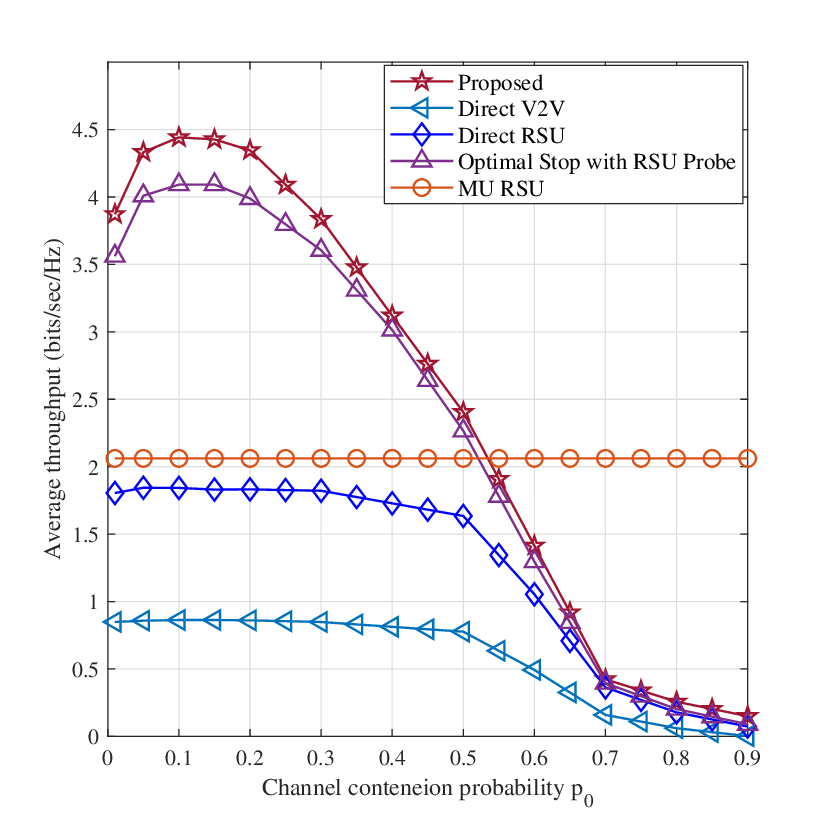}}
\caption{Average network throughput comparison of alternative strategies.}\label{f:fig7}
\vspace{-5mm}
\end{figure*}
We compare our algorithm with four existing strategies:
i) Direct V2V Strategy: User pairs transmit data directly after channel contention; ii)  Direct RSU Strategy: User pairs probe the RSU \cite{Kimura2021}; iii) Optimal Stop Strategy with RSU Probe: User pairs probe the RSU and use the optimal stop strategy for channel access \cite{Wei2020acm}; and iv) MU RSU Strategy: Multiple sources access the channel within a coherent duration. 
Adaptive rates are used to mitigate interference   \cite{Kimura2021}.
In the mobility scenario, all optimal stop strategies are updated every large-scale phase.

Fig.~\ref{f:compared_b3} compares the average throughput against transmit power $P_s$ for the proposed RPCA strategy and four alternatives, with $P_r=P_s$ and $\tau_d=15$ ms. The proposed RPCA strategy consistently outperforms all alternatives across the range of $P_s$. At $P_s=24$ dBm, it achieves significant gains: 407\% higher throughput than the direct V2V strategy, 129\% higher than the direct RSU strategy, 8.4\% higher than the optimal stop strategy with RSU probe, and a remarkable 79.6\% higher than the MU RSU strategy.

Fig.~\ref{f:move_taud_compare} shows average network throughput against channel coherence time at $P_s=P_r=26$\,dBm. The proposed distributed MAC outperforms all alternatives consistently across $\tau_d$. At $\tau_d=7$\,ms, it achieves over 105\% higher throughput than direct strategies, 248\% more than direct V2V, 8.9\% more than MU RSU, and 14.8\% more than the optimal stop strategy with RSU probe. 
Fig.~\ref{f:compare_p0} shows the average network throughput with varying channel contention probability \(p_0\) from 0.01 to 0.9, while keeping \(P_s=P_r=26\) dBm, \(\tau_d=15\) ms. 
The proposed MAC consistently outperforms the alternatives, with a significant throughput increase when \(p_0 < 0.5\). For \(p_0 \ge 0.5\), the proposed MAC approaches the performance of two alternatives using an optimal stop strategy due to increased channel contention overhead. 

\vspace{-2mm}
\section{Conclusion}\label{s:con}\vspace{-1mm}
This research addresses the challenge of distributed cooperative MAC in VANETs with the RSU.
Utilizing the theory of optimal sequential planned decision, we {have developed} the RSU probing and cooperative access (RPCA) strategy, achieving rigorous network throughput maximization. The proposed optimal strategy and its simplified forms exhibit complexities of $\mathcal{O}(1)$, respectively. Building on the RPCA strategy, we {develop} a novel location-aware MAC algorithm with two time-scale phases for distributed network operations. 
The theoretical framework, optimal RPCA strategy, and location-based implementation algorithm contribute to improved performance and reliability. 
This research also sets the stage for future advancements in self-adaptive TDMA-based MAC protocols, supporting diverse high-priority applications. 

\end{document}